\documentclass[useAMS,usenatbib,12pt,epsfig]{article}
\usepackage[english]{babel}
\usepackage{amssymb}
\usepackage{graphicx}

\begin{document}
\selectlanguage{english}

\thispagestyle{empty}

\bigskip
\bigskip
\bigskip
\begin{center}
\begin{Large}
{Some Possible Solution of Problem of Sovereign Debts: \\a short plan}
\end{Large}
\end{center}
\bigskip
\bigskip
\bigskip
\begin{center}
{Kholupenko T.S.$^*$\footnote{tanyaspb@mail.ru}, Kholupenko E.E.$^{**}$\footnote{eug\_khol@mail.ru}, Guseva P.A.$^{***}$}
\end{center}
{$^{\rm *}$Nevskaya Concession Company, St.-Petersburg 195027, Russia}
\\{$^{\rm **}$Ioffe Physical-Technical Institute, St.-Petersburg 194021, Russia}
\\{$^{\rm **}$Bank ``Russian Financial Corporation'', Moscow 125009, Russia}
\bigskip
\bigskip
\bigskip
\bigskip
\begin{center}
\bf{Abstract}
\end{center}
{
Possible solution of problem of sovereign debts is suggested. 
At the current moment this solution still can be provided only by methods of the world monetary policy. 
}
\bigskip
%\\{PACS numbers: }

\newpage
\section{Introduction}
\hspace{1.1cm}
History teaches us that economic and financial crisises are inseparable part of evolution of economy and free markets. 
Number of deep crisises such as Tulipomania Crash 1637 or Great Depression 1929 makes us to think that similar crisises will repeat 
despite any efforts to overcome and/or to prevent them. So one can be sure that we will face financial crisises as long as free financial 
markets exist.

Maybe economic crisises are necessary to stimulate subsequent development of real sector through significant drop of 
cost of labour force and resources, and cleaning of financial sector through bankruptcies and crashes 
of unreliable financial companies. Maybe crisises are product of our overestimated expectations (Sornette and Woodard 2009, Sornette and Cauwels 2012).
Whether crisises are objective factors or not, whether they are useful for economy on large time scales or not, 
but their influence on all sectors of economy during their active stage 
(beginning, crash and possible subsequent recession) is negative mostly.  That is why we should develop 
the methods of forecast of crisises (e.g. Lagi and Bar-Yam 2012) and fast overcoming of their negative consequence. 

In this paper one more possible method of overcoming of global financial crisis (which has started in 2008) is suggested. 
The main aim of suggested plan is to save the market of sovereign debts (i.e. to prevent the total crash of it). We focused on saving sovereign debts because 
of the following reason: the citizens of debtor-countries may not be guilty for huge government debt (because they are not responsible for any 
decisions about debt, even if they use amenities obtained due to this debt) 
but they will feel all horrible consequences of government default in the case of it. This differs from the situation with non-government 
(i.e. private) debts when those people who make decisions will be responsible as it should be. Also the governments make a medium for doing business. 
Thus the maximal trust to the governments and their possibilities to control economic and financial situation should be one of the main aims of efforts of 
economists.

The plan or some points of it maybe useful only if they will be applied by group of countries with leading economies: 
(G7+BRICS), G20 or even UN. 
Also, the proposed efforts should be made as soon as possible, before the onset of the second wave of the global crisis (e.g. subsequent government defaults (Lagi and Bar-Yam 2012)) 
and separate efforts of governments for local overcoming of crisis in their countries with unpredictable consequences for 
global economy (e.g. nationalization of assets of foreign residents, or uncontrolled hidden mining of electron currencies such as BITCOIN (Satoshi 2008) 
or anonymous emission of new electron currencies, and others) will become true. 

\section{Solution of Problem of Sovereign Debts}
Key idea of suggested plan is the synergy of three well-known points:
\\1) Creation of New Special World Bank (hereinafter NSWB) as an financial regulator 
\\2) Emission of New World Currency (hereinafter NWC) by NSWB
\\3) Transfer of sovereign debts into NSWB which will become the largest "dump" bank to collect bad debts.

For successful operating of this new financial institute the following conditions should be satisfied:
\\1) NSWB emits new world currency NWC (like special drawing rights [SDR, code XDR] by IMF).
 In the difference with XDR, transfer ruble or gold franc, this new currency should be currency in full meaning:

\hspace{0.2 cm}
a) NWC should circulate freely in the whole world (or in the countries which are founders of NSWB as minimum 
[this should be guaranteed by Agreement among them]). 

\hspace{0.2 cm}
b) NWC should be accessible for the states (in the person of Central Banks) as well as for commercial companies and natural persons,
 i.e. for all who can carry out financial operations (in the difference with XDR for example). 
Cash and cashless payments should be accessible with NWC.  This is very important to make significant 
fraction of cash NWC ($M0$ should be non less than 10\% of total quantity of NWC $L$) 
because cash payments are most simple type of payments. In the difference with cashless payments or 
electron currency using, cash payments are accessible for any persons, especially for natural persons (e.g. for retail buying) independently with their 
level of revenue or education. That provides very fast circulation of cash money in comparison with other types of funds. In considered case 
this actually will provide the backing of NWC by goods, commodities and resources in the whole world (this situation is similar to the situation with the actual backing of USD).
\\2) The total quantity $L$ of NWC should cover the total debt of debtor-countries. 
%(it is desirable to have multiple covering of total value of debts)
\\3) Let $a_{ik}$ denotes credit of creditor $i$, given to debtor-country $k$.

If creditor wants, part of credit should be discharged immediately by NSWB using NWC at the size of $c_{ik}\le a_{ik}$. 
Total quantity of undischarged parts of credits of creditor $i$ is given by the following expression: 
\begin{equation}
A_{i}=\sum_{k}\left(a_{ik}-c_{ik}\right)
\end{equation}

This residual credit portfolio $A_{i}$ should be converted to the deposit in NSWB with new rate $\delta^A$ reconsidered to decrease in comparison with 
current rates of credits.
\\4) Let $b_{ki}$ denotes debt of debtor-country $k$ to creditor $i$ ($b_{ki} = a_{ik}$).
Part of debt $b_{ki}$ of debtor-country $k$ should be canceled at the size $c_{ik}$ corresponding to outpayment of NSWB to creditor $i$. This 
cancellation should be performed without any obligations puted on debtor-country. 
Uncanceled part of debts of debtor-country $k$ is given by the following expression:
\begin{equation}
B_{k}=\sum_{i}\left(b_{ki}-c_{ik}\right)
\end{equation}
This residual total debt $B_{k}$ should be converted to the loan of NSWB to  debtor-country $k$ with new rate $\delta^{B}_{k}$.
The value $\delta^{B}_{k}$ depends on reliability of debtor-country $k$. Certainly $\delta^{B}_{k}$ is significantly larger than $\delta^A$ but reasonably 
less than current credit rates.

{\bf Operations of immediate discharging 3) and cancellation 4) are the one-time actions and should not be repeated in the frame of suggested plan.} 
Determination of values $c_{ik}$ is based on desire and right of creditors to define size of part of their funds (distributed previously in sovereign obligations) 
that they want to convert into new reliable deposit in NSWB with fixed deposit rate $\delta^A$.
\\5) Free market of sovereign debts should be canceled: all new loans should be made via NSWB and denominated in NWC.
 Another scenario: Free market of sovereign debts should not be regulated by International Laws but only two-side contracts. 
So they will not be guaranteed by NSWB or any other International Financial Institutions. This will make such a loans very risky.
\\6) Exchange rates of NWC to other currencies should be stated by only NSWB.
\\7) In addition to free circulation in the whole world, NWC should also be free convertible in the whole world.
%but with some restriction: exchange rate should not be more profitable than exchange rate of NSWB.
\\8) It is necessary to open enough number of offices of NSWB for service of natural and juridical persons in the whole world.

Maybe it will be useful to make some additional efforts for increasing reliability of NSWB and NWC, for example:
\\1. To use NWC as reserve currency, i.e. to oblige Central Banks of founder-countries to include some fraction of NWC in their monetary reserves.
\\2. To provide the backing of some part of NWC by means of creation of proper additional monetary reserves of NSWB in national currencies such as USD, EUR, JPY, and others. 
\\3. It is possible to provide the backing of NWC in some specific way, e.g. by some not very liquid resources of founder-countries (e.g. pure water, or not very liquid available lands). 
This backing should be regulated by Agreement.
\\4. Maybe it is reasonable to set the rate of emission of NWC as a function of time (similar to the predictable emission of BITCOIN) based on prediction of evolution of gross world product (GWP).

It should be noted that some similar steps have already been suggested and very well-known. For example the introducing new world currency has been suggested by R. Mundell for 
increasing international financial discipline. The main difference of present proposal from proposal by Mundell is the aim of using this new world currency. It seems that introducing NWC may 
give us unique possibility to cancel sovereign debts (or any desired part of them) immediately. 
Maybe it is possible to run this process without negative consequences such as inflation of national 
currencies because of absence of explicit connection of NWC with national currencies and national GDPs. 

\section{Preconditions of realization and main expected results}
Creation of this system may be supported by world financial players: 
\\a) China wants to create new world currency which would not be connected with state of US economy strongly. 
Funds obtained by China as result of discharging debt can be used for solution of some internal problems, 
e.g. stimulating domestic demands by issue of consumer credits (also in NWC) in a volumes which were not 
accessible earlier, creating pensionary system, and others.
\\b) Another creditors (e.g. Japan) can also resolve some internal problems: stimulating domestic demands 
by issue of consumer credits in NWC that may allow them to avoid significant increase of internal inflation.
\\c) Using new suggested system the large debtors (e.g. USA, EU, etc.) will be able to resolve their problems with sovereign 
debts using NWC without any threat of national defaults, necessity of devaluation of national currencies and undermining the trust to their economies.

Realization of suggested plan will allow us to resolve two key problems of modern world financial system: 
\\1) To save the market of sovereign debts from the current world crisis; 
\\2) To create new effective world currency. Also NWC will allow us to resolve Triffin Dilemma, because NWC 
will not be a national currency at all, in the difference with any other existing world currency (excluding XDR which is not currency truly).

In addition, by creation of NSWB as financial regulator and only holder of sovereign debts we can increase financial discipline in 
the world scale and try to avoid similar crisises in near future. 

\section{Obstacles for realization}
It is completely clear that many current large government debtors are ``too big to fail''. 
This is very convenient situation for these debtors. Thus realization of proposed plan depends on 
answers on following principal questions: 
\\1) Do creditors really want to return their money?
\\2) Do debtor-countries really want to pay off debt?
\\3) Are the countries ready to lose a part of their sovereignty by introducing free circulation of NWC?

\section{References}
Lagi M. and Bar-Yam Y., 2012, arXiv:1209.6369v1
\\Mundell R., http://robertmundell.net/economic-policies/world-currency/
\\Satoshi N., 2008, http://bitcoin.org/bitcoin.pdf
\\Sornette D. and Woodard R., 2009, arXiv:0905.0220v1
\\Sornette D. and Cauwels P., Notenstein Academy White Paper Series, 2012, arXiv:1212.2833
\end{document}